# Surfactant chain length and concentration influence on the interfacial tension of two immiscible model liquids: a coarse – grained approach

R. Catarino Centeno[1,2], R. A. Bustamante – Rendón[2], J. S. Hernández – Fragoso[3], I. Arroyo – Ordoñez[2], E. Pérez[3,4], S. J. Alas[3], A. Gama Goicochea[2†]

[1]Posgrado en Ciencias Aplicadas, Facultad de Ciencias, Universidad Autónoma de San Luis Potosí, Lateral Av. Salvador Nava s/n, 78290 San Luis Potosí, Mexico

[2]Posgrado en Ingeniería Química, División de Ingeniería Química y Bioquímica, Tecnológico de Estudios Superiores de Ecatepec, Av. Tecnológico, 55210, Estado de México, Mexico

[3]Posgrado en Ciencias Naturales e Ingeniería, Departamento de Ciencias Naturales, Universidad Autónoma Metropolitana Unidad Cuajimalpa, Av. Vasco de Quiroga 4871, 05300 Ciudad de México, Mexico

[4]Instituto de Física, Universidad Autónoma de San Luis Potosí
Álvaro Obregón 64, 78000 San Luis Potosí, Mexico

## Abstract

The interfacial tension between immiscible liquids is studied as a function of a model linear surfactant length and concentration using coarse–grained, dissipative particle dynamics numerical simulations. The adsorption isotherms obtained from the simulations are found to be in agreement with Langmuir´s model. The reduction of the interfacial tension with increasing surfactant concentration is found to display some common characteristics for all the values of chain length modeled, with our predictions being in agreement with Szyszkowski's equation. Lastly, the critical micelle concentration is predicted for all surfactant lengths, finding exponentially decaying behavior, in agreement with Kleven's model. It is argued that these findings can be helpful guiding tools in the interpretation of available experiments and in the design of new ones with new surfactants and polymers.

**Keywords**: Interfacial tension, surfactant chain length, adsorption, critical micelle concentration, dissipative particle dynamics.

† Corresponding author. Email: agama@alumni.stanford.edu

# INTRODUCTION

The study of surfactants' properties and structure and their connection with surfactants ability to reduce the interfacial tension efficiently are subjects of considerable current interest due to their applications to topics that range from nanomedicine to enhanced oil recovery. A surfactant molecule is an active agent which is characterized by its adsorption characteristics and capacity to reduce interfacial tension [1]. Most surfactant molecules consist of at least two parts: a hydrophilic (polar) head and a hydrophobic (non-polar) tail, which possess the property of reducing interfacial tension in an aqueous medium.

The self-assembling of surfactants in solution is a phenomenon present in natural systems and industrial processes, such as pharmaceuticals, food, plastics and petroleum industries. In recent years research in this area has played an important role in the science of colloids and interfaces [2]. Since surfactants naturally adsorb at interfaces, one of their physicochemical characteristics is that they can saturate interfaces when they reach certain concentration, known as the critical micellar concentration (CMC). When the surfactant concentration exceeds the CMC, the surfactants aggregate in micelles, which differ in size and shape and coalesce more as the surfactant concentration increases even further [3]. Knowing precisely the CMC is important from a practical point of view, since adding surfactants at concentrations above CMC can be wasteful and costly. Also, one would like to know how the CMC and the interfacial tension depend on the surfactant's particular structure. Numerical modeling can be a very powerful tool to tackle these and other related problems [4, 5]. In particular, surfactants micellization has been extensively studied since the 1990's using computer simulations [6-12]. However, many surfactant applications do not require atomistically detailed simulations [4], which can be computationally demanding if one wants to reach time and length scales of the order of ~ 1 μs and ~ 100 nm, respectively, as is customary in the surfactant consuming industries [1]. Reaching these scales can be achieved using so called coarse – grained methods, for example [13].

Some years ago Hoogerbrugge and Koelman [14] introduced a new simulation technique called dissipative particle dynamics (DPD). It is based on the simulation of soft spheres ("beads"), whose motion is governed by simple force laws; in addition, it allows for the mesoscopic-scale modeling of the self-assembly of surfactant and polymer systems. DPD is based on a coarse – grained representation (CG), where the internal degrees of freedom of the molecules are integrated out in favor of a less atomistically detailed and more mesoscopic description of the system. Beads interact through soft, short range potentials that lead to improved computational efficiency. Despite the simplicity of the models, DPD can provide quantitatively and qualitatively correct descriptions of structural and thermodynamic properties of complex systems [15, 16].

One of the main problems to overcome to achieve quantitatively correct predictions using DPD is the parameterization of the repulsion parameters. Poll and Bolhuis [17] performed Monte Carlo (MC) simulations using soft repulsion potentials taken from [18] and obtained very low values for the CMC. The authors suggested that the inclusion of a hard sphere solvent is recommended to obtain quantitatively accurate CMC predictions using CG models. However, in recent publications by Li et al. [19] and by Lin et al. [20] it is reported that CMC studies of non-ionic model surfactants with different structure showed that DPD simulations predicted correctly the formation dynamics and the equilibrium distribution of micellar aggregates. The influence of varying the bond strength between beads along the surfactant chain in their ability to reduce interfacial tension and other thermodynamic properties has also been studied using DPD [21]. The role of the surfactant chain rigidity in the prediction of the CMC was explored by Lee and collaborators using DPD simulations [3]. On the experimental side, the work of Mattei et al. [22] reported the use of the group-contribution method (GCM) to estimate the CMC of nonionic surfactants with different molecular structures in water at 298.15 K; the GCM is based on Merrero and Gani's model [23]. Determining the value of the CMC experimentally can be carried out using different physical quantities such as electrical conductivity, surface tension, light scattering and fluorescence spectroscopy depending on the size or number of particles in solution [24]. Additionally, some applications require surfactant concentrations just below the CMC (e.g. for reduction of surface tension) while for others, such as in the design of emulsion – based products, it is necessary to work above the CMC [22]. Since the formation of the interface is a dynamic process, the effectiveness of the surfactants is determined by the rate at which the surfactant is adsorbed at the interface, as well as by the thermodynamic competition between adsorption and micelle formation, which determines how effectively a surfactant reduces the equilibrium surface tension [25].

In the present work we report studies of the interfacial tension between two immiscible model liquids ("water"/"oil") as a function of the model surfactant's chain length and concentration, using DPD numerical simulations. Our aim is to determine not only the role played by the chain length in reducing the interfacial tension, but also the effect of the competition between the adsorption and aggregation phenomena in the surfactants' efficiency. Five types of linear, non - ionic surfactants with increasing degrees of polymerization were modeled ($N = 30, 40, 50, 60, 70$ beads). We focus also on identifying how surfactants adsorb, so that one can establish where in concentration is the critical region of saturation of surfactants at the interface, which yields directly predictions of the CMC. The emphasis is on the study of the thermodynamic mechanisms at play rather than in the mapping to specific molecular structures, so that our conclusions can be useful to a wider number of systems.

## MODELS AND METHODS

Our models were solved numerically using the DPD method, in which the fluid is composed of a set of point particles in continuous space and the motion of the particles is governed by Newton's laws. The system consists of a set of *n* particles, where each particle is characterized by its position $r_i$, moment $p_i$, and its mass $m_i$. The evolution of positions and momenta on all particles in time is obtained from the numerical integration of Newton's second law on discreet time steps. The net force on the *i*-th particle, $F_i$ is given by

$$F_i = \sum_{i \neq j} F_{ij}^C + F_{ij}^D + F_{ij}^R \quad , \quad (1)$$

where the DPD interparticle force exerted by the particle *i* on the particle *j* is additive in pairs and is made up of three components: conservative $F_{ij}^C$, dissipative $F_{ij}^D$ and random $F_{ij}^R$. These forces are given by the following expressions:

$$F_{ij}^C = \begin{cases} a_{ij}\left(1 - \frac{r_{ij}}{r_c}\right)\hat{r}_{ij} & \frac{r_{ij}}{r_c} \leq 1 \\ 0 & \frac{r_{ij}}{r_c} > 1 \end{cases} \quad (2)$$

$$F_{ij}^D = \begin{cases} -\gamma \omega^D(r_{ij})[v_{ij} \cdot \hat{r}_{ij}]\hat{r}_{ij} & \frac{r_{ij}}{r_c} \leq 1 \\ 0 & \frac{r_{ij}}{r_c} > 1 \end{cases} \quad (3)$$

$$F_{ij}^R = \begin{cases} \sigma \omega^R(r_{ij})\xi_{ij}(\Delta t)^{-1/2}\hat{\boldsymbol{r}}_{ij} & \frac{r_{ij}}{r_c} \leq 1 \\ 0 & \frac{r_{ij}}{r_c} > 1 \end{cases} \quad (4)$$

where $\boldsymbol{r}_{ij} = \boldsymbol{r}_i - \boldsymbol{r}_j$ is …; $\boldsymbol{v}_{ij} = \boldsymbol{v}_i - \boldsymbol{v}_j$ is …; $r_{ij} = |\boldsymbol{r}_{ij}|$ and the unit vector $\hat{\boldsymbol{r}}_{ij} = \boldsymbol{r}_{ij}/r_{ij}$. $\xi_{ij}$ is a random variable with a Gaussian distribution [14], $\omega^R(r_{ij})$ is a weight function, $\gamma$ the friction factor and $\sigma$ defines the amplitude of the fluctuations; $\Delta t$ is the finite time step used to integrate the equation of motion. The constants $\sigma$ and $\gamma$ are coupled by the relation $\sigma^2 = 2\gamma k_B T$, with $k_B$ the Boltzmann constant and $T$ the system's absolute temperature, respectively. The random and dissipative forces are balanced and related to the system temperature according to the fluctuation-dissipation theorem [27]:

$$\omega^D(r_{ij}) = [\omega^R(r_{ij})]^2 = [max\{(1 - r_{ij}/r_c), 0\}]^2. \quad (5)$$

When the above relationship is fulfilled, together with that existing between $\sigma$ and $\gamma$, the system reaches thermodynamic equilibrium, having a canonical distribution function [27]. The equations of motion are solved using the velocity Verlet algorithm adapted to DPD [26]. The DPD particles interact only with those within a certain cutoff radius $r_c$ beyond which all interactions are zero. The value $r_c = 1$ sets the model's length scale. Dissipative and random forces constitute the local DPD thermostat. Surfactants are modeled by linear chains of beads connected by harmonic springs between beads. The harmonic force that joins the beads is given by:

$$\boldsymbol{F}_{spring}(r_{ij}) = -k_0(r_{ij} - r_0)\hat{\boldsymbol{r}}_{ij}, \quad (6)$$

where $k_0$ is the spring constant and $r_o$ the equilibrium distance of the spring, respectively. In all simulations the values of these constants are $k_0 = 100(k_B T/r^2_c)$ and $r_0 = 0.7r_c$ [21].

## SIMULATION DETAILS

All simulations are performed in reduced DPD units. The time step chosen to integrate the equation of motion is $\Delta t^* = 0.01$; the volume of the rectangular parallelepiped simulation box for all cases is $L_x \times L_y \times L_z = 10 \times 10 \times 30$. The values of the constants involved in the dissipative and random forces are chosen (see Eqs. (3-4)) as $\sigma = 3$ and $\gamma = 4.5$, respectively, so that $k_B T^* = 1$. The cut-off radius is $r_c = 1$, and the masses are $m_i = m = 1$. Reduced units are used throughout this work, however the coarse – graining degree ($N_m$) we use is three water molecules per DPD bead, which sets the length scale as $r_c = (\rho N_m V_m)^{1/3} = 6.46$ Å, with $\rho = 3$ being the numerical density and $V_m = 30 Å^3$ the molar volume of a water molecule. The reduced time step ($\Delta t^*$) is dimensionalized with $\Delta t = (mr_c^2/k_B T)^{1/2} \Delta t^* \approx (6.3\ ps)\ \Delta t^*$ for $N_m = 3$; the interfacial tension is expressed in units of $k_B T/r_c^2$ and the interaction parameters $a_{ij}$ are given in units of $k_B T/r_c$. There is a well-established method for obtaining the interaction parameters of the conservative force between the various types of particles in DPD using the Flory – Huggins solution theory, see reference [28]. However, our purpose here is to show how DPD can be used to extract important information about surfactant efficiency quickly and accurately, therefore the values for the conservative forces are chosen heuristically for interactions between particles of different type. For like – like interactions the conservative force constant is given by the coarse – graining degree and the reduced value of the isothermal compressibility of water [29]. Since our coarse – graining degree is equal to grouping three water molecules into a DPD bead, the appropriate value for the conservative force parameter (see Eq. (3)) for particles of the same type is $a_{ii} = 78.3$, in reduced DPD units [21]. The interactions for different types for particles were chosen qualitatively, to produce strong repulsion (as for

water – oil, and water – surfactant tail interactions) and strong attraction (as for water – surfactant head, and oil – surfactant tail interactions), where experience dictates that they are so. For example, since the diagonal interactions in Table 1 are close to 80 in reduced DPD units, an interaction twice as large must produce strong repulsion.

The simulations were performed in blocks of $2 \times 10^4$ time steps, with the first 20 blocks being used to reach equilibrium and the additional 20 blocks used for the production phase. All numerical calculations were performed in the canonical ensemble (at constant particle number, $n$, volume, $V$, and temperature, $T$), for a mixture of two monomeric immiscible liquids and with linear chain surfactants at the interface with increasing degree of polymerization ($N$ = 30, 40, 50, 60, 70). All simulations are carried out at fixed global density $\rho^* = 3$, which means that the total number of beads (water, oil, surfactants) in the system is always $n = 9000$, not to be confused with the surfactant's polymerization degree. Periodic boundary conditions are applied on all faces of the simulation box.

## RESULTS AND DISCUSSION

The surfactants modeled in this work have a single – monomer hydrophilic head and a hydrophobic tail of increasing length, immersed in an aqueous medium formed by two immiscible components, water and oil, forming interfaces. Table 1 lists the values of the interaction parameters used for all simulations, see Eq. (2).

**Table 1**. Conservative force interaction constants between the beads. Here 1 = water, 2 = oil, 3 = surfactant's polar head and 4 = surfactant's hydrophobic tail. See also Eq. (2).

| $a_{ij}$ | 1 | 2 | 3 | 4 |
|---|---|---|---|---|
| 1 | 78.3 | 160 | 60 | 160 |

| | | | |
|---|---|---|---|
| 2 | | 78.3 | 160 | 60 |
| 3 | | | 78.3 | 85 |
| 4 | | | | 78.3 |

Figure 1(a) shows a snapshot of our model system without surfactants (left image), along with their respective density profiles (right image). On the left side in Fig. 1(b) we present a snapshot of the water-oil-surfactant system and the corresponding density profiles (right panel). The linear surfactant presented in this figure corresponds to the case of degree of polymerization equal to 30.

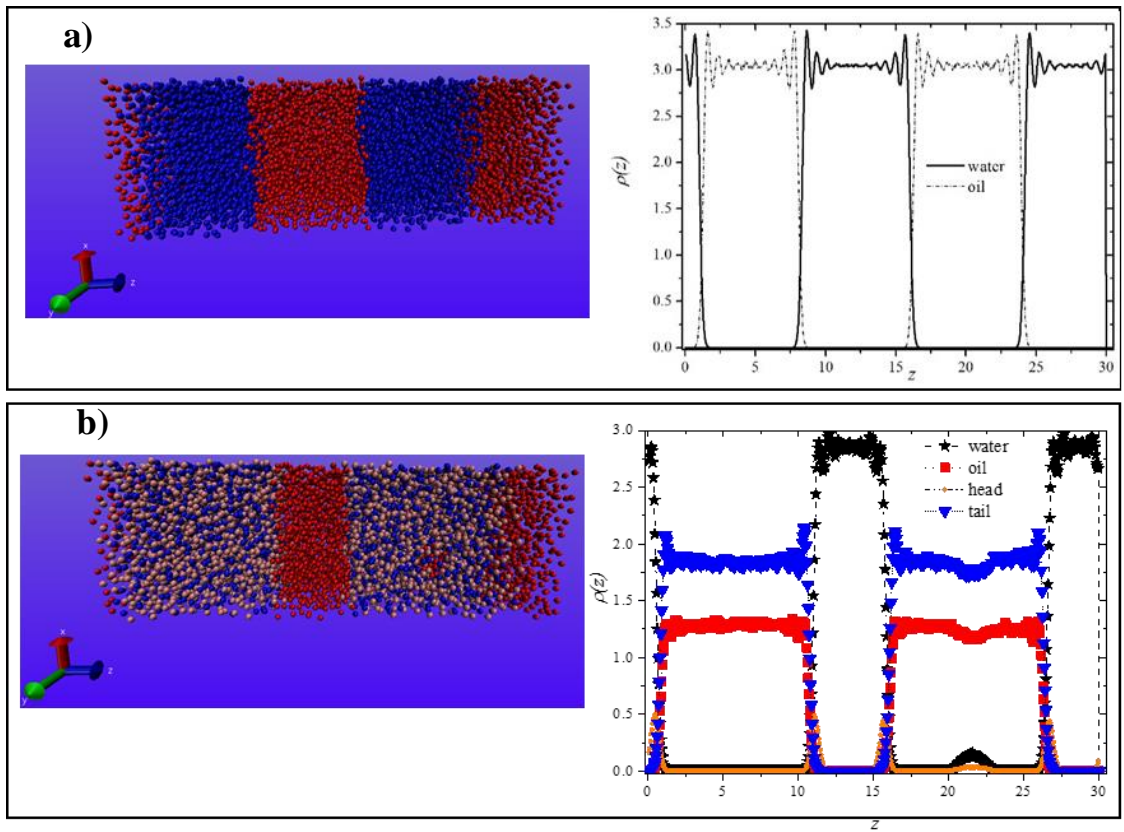

**Figure 1**. (Color online) (a) Snapshot of the water (red beads) – oil (blue beads) system (left) and their respective density profiles, on the right (solid line for water, dash-dot line for oil). (b) Snapshot of the water-oil-surfactant system (left), and their corresponding density profiles (right; orange circles for the hydrophilic head, blue

triangles for the hydrophobic tail, black stars for water and red circles for oil; $N = 30$) at a concentration of 130 chains/vol. All quantities are expressed in educed DPD units. The snapshots of the systems were obtained with the molecular visualizer VMD [30].

The density profiles shown in Fig. 1 display the formation of four interfaces due to periodic boundary conditions. Notice also how the profiles in Fig. 1(b) show the incipient formation of a surfactant micelle within the oil phase, in addition to the adsorption of surfactants at the interfaces.

The interfacial tension $\gamma^*$ of the system is obtained from the components of the pressure tensor, of the form

$$\gamma^* = L_z^*\left[\langle P_{zz}^*\rangle - \frac{1}{2}\left(\langle P_{xx}^*\rangle + \langle P_{yy}^*\rangle\right)\right], \qquad (7)$$

where $\langle \cdots \rangle$ is the time average of the components of the pressure tensor over the production phase of the simulations, $L_z^*$ is the length of the simulation cell along the $z$ – direction, which is perpendicular to the interfaces shown in the snapshots in Fig. 1. The asterisks indicate expressions are given in reduced units. The components of the pressure tensor are obtained from the virial theorem [4], as follows:

$$P_{xx} = \sum_i m_i \vec{v}_i \cdot \vec{v}_i + \sum_i \sum_{j>i} F_{ijx}^C x_{ij}. \qquad (8)$$

The first term in Eq. (8) represents the kinetic contribution, while the second term arises from the interactions and is the product of the $x$ – component of the conservative DPD force acting between particles $i$ and $j$, and the $x$ – component of the relative position vector between particles $i$ and $j$. The other pressure tensor components, $P_{yy}$ and $P_{zz}$ are obtained by replacing $x$ by $y$ and $z$ in Eq. (8), respectively. It has been shown [31] that only the conservative force contributes to the pressure tensor for long simulations, therefore the dissipative and random forces are not included in Eq. (8) [28].

Figures 2(a) and 2(b) show the adsorption isotherms for linear surfactants with a degree of polymerization equal to 30 and 50 units, respectively. These isotherms represent the concentration of surfactant adsorbed at the interfaces versus the concentration of the surfactant not adsorbed (forming micelles), also known as the supernatant; in both cases the data are compared with the Langmuir adsorption isotherm [32].

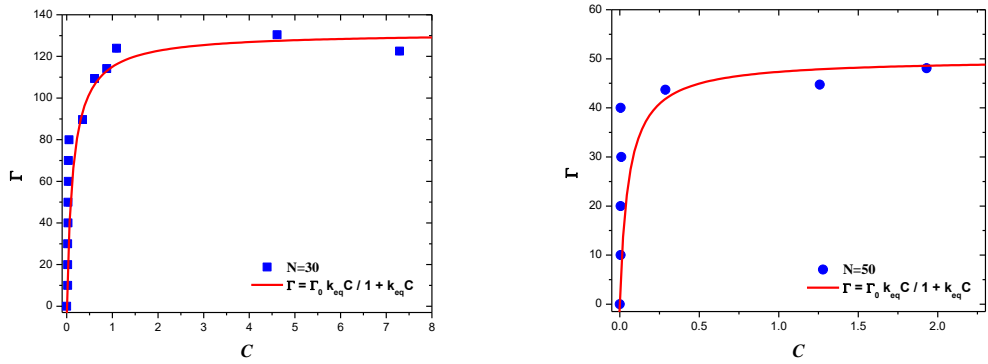

**Figure 2**. (Color online) (a) Adsorption isotherm for the surfactant with polymerization degree $N = 30$ and (b) adsorption isotherm for the surfactant with $N = 50$, in reduced DPD units. The solid lines represent the fit to the Langmuir adsorption model, eq. (9) with $k_{eq}(N = 30) = 7$ and $k_{eq}(N = 50) = 20$, also in reduced units. The error bars are smaller than the symbols' size.

The trends in the data in Fig. 2 are shown to be adequately reproduced by the Langmuir model [32],

$$\Gamma = \Gamma_0 \frac{k_{eq}C}{1+k_{eq}C}, \qquad (9)$$

where $\Gamma$ is the number of chains adsorbed at the interfaces per unit area, $\Gamma_0$ is the adsorption at saturation, $k_{eq}$ is the equilibrium constant and $C$ the surfactant concentration. The agreement between our predictions and Eq. (9) is the result of the fact that the surfactant adsorbs at the interfaces primarily as a monolayer, as the chains' heads profile in Fig. 1(b) shows. The fits in Fig. 2 show that $k_{eq}(N = 30) < k_{eq}(N = 50)$, as has been found in various experiments [1].

In Fig. 3 we present the dimensionless interfacial tension as a function of surfactant concentration, normalized by their corresponding CMC, for increasing values of the chains' length. The CMC is obtained from the concentration at which there is an abrupt change in the slope of tension, as the inset in Fig. 3 illustrates. The interfacial tension decreases as the surfactant concentration increases, as expected for any surfactant and found in numerous experiments [22, 25] and in simulations [3, 6, 21]. The simplest model that is adequate to describe the surfactant adsorption is the Langmuir isotherm [33], which reproduces our results reasonably well, as seen in Fig. 2. This means that the adsorption-desorption kinetics of the surfactant and its diffusion in the system determines the rate of adsorption of the surfactant [25].

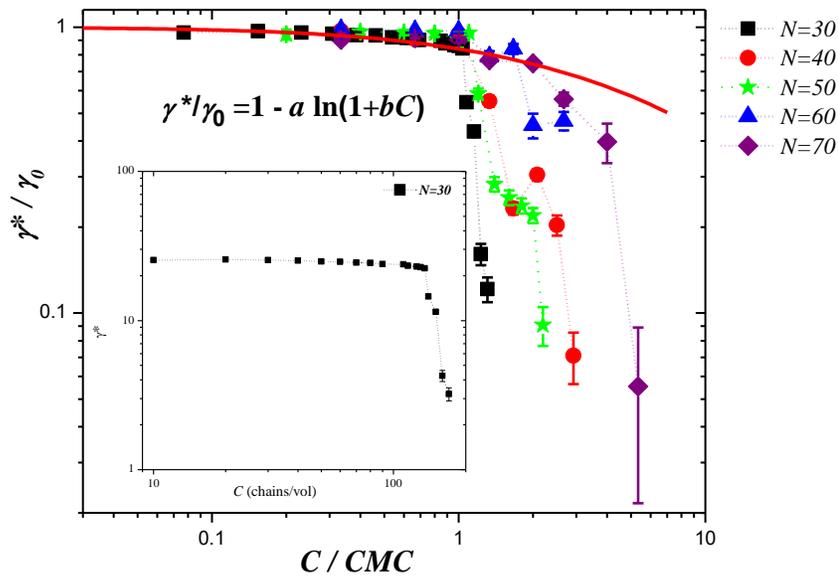

**Figure 3**. (Color online) Normalized interfacial tension as a function of the variation of the concentration (normalized by the CMC) of the model surfactants for increasing chain length, in reduced units. The solid line represents the fit to Syzckowski's equation [25], with $a = 0.25$ and $b = 0.9$. The inset shows the interfacial tension as a function of the variation of the concentration of the surfactant with polymerization degree $N = 30$, in reduced DPD units.

The data for all surfactants fall onto a single curve for concentrations below their CMC, as seen in Fig. 3, which signals scaling behavior as is frequently the case for self – similar molecules like surfactants and polymers. The line in Fig. 3 is the fit to Szyszkowski's equation [25]:

$$\frac{\gamma}{\gamma_0} = 1 - a \ln(1 + bC) , \qquad (10)$$

where $\gamma_0$ is the interfacial tension when no surfactants are present in the system, while $a = RT\Gamma_\infty/\gamma_0$ and $b = (k_{eq})(CMC)$ are constants that are usually obtained from fits to experimental adsorption and surface tension isotherms; $R$ is the universal gas constant and $\Gamma_\infty$ is the adsorption at maximum saturation of the interface. No attempt was made to fit the data in Fig. 3 for $C > CMC$ since that would fall beyond the range of validity of Eq. (10). Therefore the solid line in the main panel in Fig. 3 is the fit for the normalized interfacial tension data of all surfactant lengths $N$ in the concentration range $0 < C < CMC$. Equation (10) is based on Langmuir's isotherm and Gibbs adsorption equation, therefore the agreement between Szyszkowski's equation and our predictions in Fig. 3 is a consequence of the adsorption mechanism. A salient feature of the curves shown in Fig. 3 for surfactant concentration above the CMC and their fit to Eq. (10) is the following. Increasing the chain's length $N$ reduces the CMC, which would lead to a decrease in the parameter $b$ in Eq. (10). However, a larger $N$ also leads to a larger $k_{eq}$, as Fig. 2 illustrates, which would also produce an increase in $b$; yet the trends shown in Fig. 3 indicate that $b$ decreases as $N$ growths. Therefore one must conclude that the CMC decreases faster with increasing $N$ than the increase of $k_{eq}$ with growing chain length. As Fig. 4 shows, that is precisely what our DPD simulations predict.

In Fig. 4 we present the CMC obtained from the interfacial tension isotherms for each chain length, such as the one shown in the inset of Fig. 3, as a function of the surfactant's chain length. The CMC is predicted to decrease exponentially as the surfactant's polymerization degree growths, in excellent agreement with experiments with polypropylene oxide polymers acting as surfactants at the air/liquid interface [34].

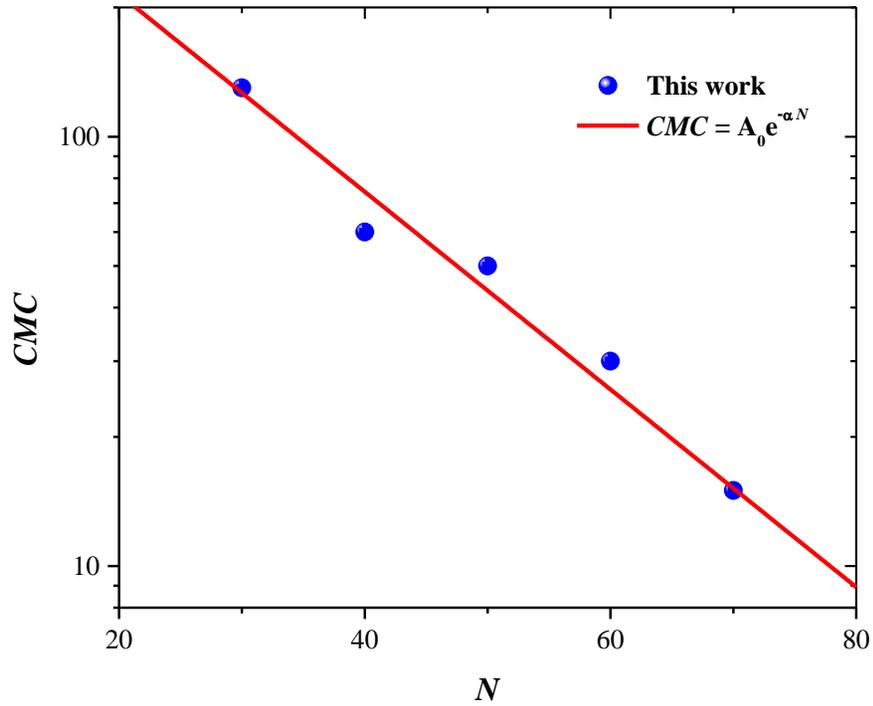

**Figure 4**. (Color online) The relationship between the CMC in reduced units predicted by DPD simulations and the length of the surfactant chain. The solid red line represents the fit to the expected exponential decay, see Eq. (11). Error bars are smaller than the symbols' size.

The solid line in Fig. 4 is the fit to the exponentially decaying dependence on polymerization degree of the CMC, expected from the model put forward by Klevens [35]:

$$\ln(CMC) = A - BN \,. \tag{11}$$

In Eq. (11) *A* and *B* are constants with respect to the surfactant chain's length, *N*. The predictions of our DPD simulations are in good agreement with experiments on several classes of non – ionic surfactants and models [34, 36, 37, 38], with the additional advantage that DPD simulations are very fast and they are easily adapted to model other surfactant architectures, such as those of the so –called Gemini [11, 37], or star – shaped [36] surfactants. If a detailed DPD mapping of a surfactant with a specific chemical composition is carried out one can obtain detailed thermodynamics information from data such as those presented in Fig. 4, e. g. the van der Waals interaction per $CH_2$ along the hydrocarbon chain, as the leading mechanism driving the micelle formation.

## CONCLUSIONS

The interfacial tension between two model immiscible liquids as a function of surfactants' concentration and chain length was calculated using mesoscopic – scale, DPD simulations. The mechanism through which the model linear surfactant decreases the interfacial tension was found to be its adsorption as a monolayer at the interfaces. The predicted adsorption isotherms were found to be in agreement with the Langmuir adsorption model, while the interfacial tension and CMC were found to behave as predicted by known theories. Although emphasis was placed on the study of model systems rather than on specific molecules and liquids, our parametrization of the surfactant has been shown to be an excellent model for polyethylene glycol in aqueous media [39], to mention a particular example. Our conclusions can be equally useful for other non – ionic surfactants [40]. Other variables can also be explored which are computationally inexpensive with this mesoscopic technique, such as the influence of larger surfactant heads on interfacial tension and on the CMC; mixtures of linear and branched surfactants in polymeric liquids, and somewhat more costly simulations, for ionic surfactants [41]. In addition to its mesoscopic reach and its speed compared with other numerical modeling techniques, DPD has also the advantage of being relatively insensitive to finite size effects, particularly in regard to the prediction of the interfacial tension [16]. Hence, one can perform accurate simulations with relatively small systems. The influence of different solvents can also be incorporated [42] through the judicious choice of conservative interaction parameters. Lastly but no less importantly, one can also carry out studies of the influence of temperature on the interfacial tension modified by specific surfactants by incorporating the temperature evolution of the conservative force parameters [43]. It is noted that through fast DPD simulations like those reported here one can gain significant knowledge by designing classes of surfactants and polymers in solution before embarking in costly and time – consuming experiments. Our main purpose here has been to show how one can use state of the art mesoscale models solved numerically to obtain useful physicochemical information that can help interpret and design better experiments on surfactant systems.

## ACKNOWLEDGEMENTS

The authors are grateful to J. Cervantes and M. A. Vaca for encouragement and support, and to J. A. Arcos, M. A. Balderas Altamirano and J. Klapp for fruitful conversations. RCC and JSHF thank CONACYT for support; RABR thanks also CONACYT (grant 252320-25985-1282). J. Limón (IFUASLP) is acknowledged for technical support. For computational resources the authors thank ABACUS, where most calculations were carried out; the high performance cluster Yoltla at UAM – Iztapalapa; Universidad de Sonora for access to the Ocotillo cluster at their High Performance Computational Area; and the Laboratorio Nacional de Caracterización de Propiedades Fisicoquímicas y Estructura Molecular Supercómputo Universidad de Guanajuato. AGG acknowledges also the computer resources and support provided by the Laboratorio Nacional de Supercómputo del Sureste de México, CONACYT network of national laboratories.

**REFERENCES**

[1] K. Holmberg, B. Jönsson, B. Kronberg, and B. Lindman. *Surfactants and Polymers in Aqueous Solution.* 2002.

[2] H. Wu, J. Xu, X. He, Y. Zhao and H. Wen. Mesoscopic simulation of self-assembly in surfactant oligomers by dissipative particle dynamics. *Colloids and Surfaces A: Physicochem. Eng. Aspects,* **290**, 239–246 (2006).

[3] M. T. Lee, V. Aleksey, and A. V. Neimark. Calculations of Critical Micelle Concentration by Dissipative Particle Dynamics Simulations: The Role of Chain Rigidity. *The Journal of Physical Chemistry B*, **117**, 10304−10310 (2013).

[4] M. P. Allen and D. J. Tildesley. Computer Simulation of Liquids, Oxford University Press, Oxford, (1987).

[5] D. Frenkel and B. Smit, Understanding Molecular Simulation, 2nd ed. Academic, New York, (2002).

[6] B. Smit, P. A. J. Hilbers, N. M. van Os, L. A. M. Rupert and I. Szleifer. Computer Simulations of Surfactant Self-Assembly. *Langmuir,* **9**, 9−11 (1993).

[7] R. G. Larson, Monte-Carlo Simulation of Microstructural Transitions in Surfactant Systems. *J. Chem. Phys.* **96**, 7904−7918 (1992).

[8] A. T. Bernardes, V. B. Henriques and P. M. Bisch. Monte Carlo simulation of a lattice model for micelle formation. *J. Chem. Phys.* **101**, 645−650 (1994).

[9] A. Jusufi and A. Z. Panagiotopoulos. Explicit and Implicit Solvent Simulations of Micellization in Surfactant Solutions. *Langmuir*, **31**, 3283-3292 (2015).

[10] S. Abel, M. Waks, M. Marchi and W. Urbash. Effect of Surfactant Conformation on the Structures of Small Size Nonionic Reverse Micelles: A Molecular Dynamics Simulation Study. *Langmuir*, **22**, 22 (2006).

[11] H. Gharibi, Z. Khodadadi, S. Morteza Mousavi-Khoshdel, S. M. Hashemianzadeh and S. Javadian. Mixed micellization of Gemini and conventional surfactant in aqueous solution: A lattice Monte Carlo simulation. *Journal of Molecular Graphics and Modelling*, **53**, 221–227 (2014).

[12] A. Nikoubashman and A. Z. Panagiotopoulos. Communication: Effect of solvophobic block length on critical micelle concentration in model surfactant systems. *The Journal of Chemical Physics*, **141**, 041101 (2014).

[13] P. Carbone and A. Carlos. Coarse-grained methods for polymeric materials: enthalpy and entropy-driven models. *WIREs Comput. Mol. Sci.* **4**, 62-70 (2014).

[14] P. J. Hoogerbrugge and J. M. V. A. Koelman. Simulating Microscopic Hydrodynamic Phenomena with Dissipative Particle Dynamics. *Europhys. Lett.* **19**, 155 (1992).

[15] T. Murtola, M. Karttunen and L. Vattulainen. Systematic coarse graining from structure using internal states: Application to phospholipid/cholesterol bilayer. *The Journal of Chemical Physics*, **131**, 055101 (2009).

[16] M. A. Balderas Altamirano, E. Pérez and A. Gama Goicochea (2017). On Finite Size Effects, Ensemble Choice and Force Influence in Dissipative Particle Dynamics Simulations. C.J. Barrios Hernández et al. (Eds.): CARLA 2016, CCIS 697, *High Performance Computing* (pp. 314–328), Springer International Publishing.

[17] R. Pool and P. G Bolhuis. Can purely repulsive soft potentials predict micelle formation correctly? *Physical Chemistry Chemical Physics,* **8** (8), 941-948 (2006).

[18] R. D. Groot. Mesoscopic simulation of polymer-surfactant aggregation. *Langmuir*, **16** (19), 7493-7502 (2000).

[19] Z. Li and E. E. Dormidontova. Kinetics of Diblock Copolymer Micellization by Dissipative Particle Dynamics. *Macromolecules,* **43** (7), 3521-3531 (2010).

[20] Y. L. Lin, M. Z. Wu, Y. J. Jane Sheng and H. K. Tsao. Effects of molecular architectures and solvophobic additives on the aggregative properties of polymeric surfactants. *Journal of Chemical Physics*, **10**, *136* (2012).

[21] R. López-Rendón, M. Romero-Bastida and A. Gama Goicochea. Dependence of Thermodynamic properties of model systems on some dissipative particle dynamics parameters. *Mol. Physics,* **105:17**, 2375 - 2381 (2007).

[22] M. Mattei, K. Georgios and R. Gani. Modeling of the Critical Micelle Concentration (CMC) of Nonionic Surfactants with an Extended Group-Contribution Method. *Ind. Eng. Chem. Res.* (2013).

[23] J. Marrero and R. Gani. Group Contribution Based Estimation of Pure Component Properties. *Fluid Phase Equilibria* **183-184**, 183-208 (2001).

[24] P. Mukerjee and K. J. Mysels. Critical Micelle Concentration of Aqueous Surfactant Systems, 1971, NSRDS-NBS 36, U.S. Dept. of Commerce, Washington, DC.

[25] J. K. Ferri and K. J. Stebe. Which surfactants reduce surface tension faster? A scaling argument for diffusion-controlled adsorption. *Advances in Colloid and Interface Science*, **85**, 61-97 (2000).

[26] I. Vattulainen, M. Karttunen, G. Besold and J. M. Polson. Integration Schemes for Dissipative Particle Dynamics: From Softly Interacting Systems Towards Hybrids Models. *J. Chem. Phys*. **116**, 3967 (2002).

[27] P. Español and P. Warren, Statistical Mechanics of Dissipative Particles Dynamics. *Europhys. Lett.* **30** (4), 191-1996 (1995).

[28] R. D. Groot and P.B. Warren, Dissipative particle dynamics: Bridging the gap between atomistic and mesoscopic simulation. *J. Chem. Phys.* **107**, 4423 (1997).

[29] R. D. Groot and K. L. Rabone. Mesoscopic simulation of cell membrane damage, morphology change and rupture by nonionic surfactants. *Biophys J.* **81**, 725-736 (2001).

[30] H. William, A. Dalke and K. Schulten. VMD: Visual Molecular Dynamics. *Journal of Molecular Graphics*, 14, 33-338 (1996).

[31] A. Gama Goicochea, M. A. Balderas Altamirano, J. D. Hernández and E. Pérez. The role of the dissipative and random forces in the calculation of the pressure of simple fluids with dissipative particle dynamics. *Computer Physics Communications,* **188**, 76-81 (2014).

[32] O. Redlich and D. L. Peterson. A Useful Adsorption Isotherm, *J. Phys. Chem.* **63** (6), 1024-1024 (1959).

[33] C. H. Chang and E. I. Franses. Adsorption dynamics of surfactants at the air/water interface: a critical review of mathematical models, data, and mechanisms. *Colloids and Surfaces A: Physicochemical and Engineering Aspects.* **100**, 1-45 (1995).

[34] Z. Zhang and H. Yin. Effect of polyoxypropylene chain length on the critical micelle concentration of propylene oxide-ethylene oxide block copolymers. *Journal of Zhejiang University SCIENCE.* **6(3)**, 219-221 (2005).

[35] H. B. Klevens. Structure and Aggregation in Dilute Solutions of Surface Active Agents. *The Journal of the American Oil Chemists' Society*. 74-80 (1953).

[36] T. Yoshimura, T. Kusano, H. Iwase M. Shibayama, T. Ogawa and H. Kurata. Star-Shaped Trimeric Quaternary Ammonium Bromide Surfactants: Adsorption and Aggregation Properties. *Langmuir*, **28**, 9322-9331 (2012).

[37] T. A. Camesano and R. Nagarajan. Micelle formation and CMC of gemini surfactants: thermodynamic model. *Colloids and Surfaces A: Physicochemical and Engineering Aspects*, **167**, 165–177 (2000).

[38] K. T. Shinoda, T. Nakagawa, B. Tamamushi and T. Isemura. (1963). Colloid Surfactant. Academic Press, New York, p.58.

[39] A. Gama Goicochea. Adsorption and Disjoining Pressure Isotherms of Confined Polymers Using Dissipative Particle Dynamics. *Langmuir*, **23**, 11656-63 (2007).

[40] A. Gama Goicochea (2013). Competitive adsorption of surfactants and polymers on colloids by means of mesoscopic simulations. J. Klapp and A. Medina, *Experimental and Computational Fluid Mechanics (Environmental Science and Engineering / Environmental Engineering)* (pp. 147-155), Springer International Publishing.

[41] F. Alarcón, G. Pérez Hernández, E. Pérez and A. Gama Goicochea. Coarse-grained simulations of the salt dependence of the radius of gyration of polyelectrolytes as models for biomolecules in aqueous solution. *Biophysics of Structure and Mechanism*, **42**(9), 661-672 (2013).

[42] A. Gama Goicochea and M. Briseño. Application of molecular dynamics computer simulations to evaluate polymer - solvent interactions. *Journal of Coatings Technology and Research*, **9** (3) 279 (2012).

[43] E. Mayoral and A. Gama Goicochea. Modeling the temperature dependent interfacial tension between organic solvents and water using dissipative particle dynamics. *The Journal of Chemical Physics*, **138** (9) (2013).